# Accretion Disc Outbursts and Stability Analysis

Liza Devi *; Asish Jyoti Boruah; Biplob Sarkar
* Corresponding author
Department of Applied Sciences, Tezpur University, Napaam, Assam-784028, India
E-mail of corresponding author: app23110@tezu.ac.in

*Abstract*— Accretion disc outbursts are re-occurring events observed in various astrophysical systems, including X-ray binaries and cataclysmic variables. These outbursts are characterized by a sudden increase in luminosity due to various instabilities in the accretion disc. We need to investigate the time-dependent accretion flow models to understand the mechanisms driving these outbursts. Time-dependent models incorporate the disc's time evolution and can capture the build-up of instabilities. This review aims to give a basic overview of accretion disc outburst and stability analysis. The paper highlights the necessity of considering the hierarchy of different timescales, dynamical, viscous, and thermal, when investigating the instabilities occurring in the accretion disc. The importance and observational implications of studying these accretion disc outbursts are also discussed.

## I. Accretion Disc Outbursts: An Overview

Accretion discs are swirling discs of gases rotating differentially around a compact object, like a neutron star or a black hole. These discs are an important component of various binary systems. These systems are consist of a compact object and a companion star, including X-ray binaries and cataclysmic variables [1,2,5,6,9]. The accretion disc is fuelled by matter drawn from the companion star and accreted onto the compact object. Sometimes, accretion discs experience a remarkable characteristic by a quick increase in brightness, followed by a slower decline. This is known as an accretion disc outburst [1,2,8]. Due to the sudden increase in the accretion rate, outbursts occur and result in a dramatic increase in luminosity. This phenomenon is often observed in X-ray binaries and cataclysmic variables. After an outburst, the system typically enters to an extended period of inactivity, known as quiescence [1]. This state can last for a significant amount of time before occurring another outburst. The period and recurrence time of outbursts vary from system to system ranging from minutes to months and even years.

This review aims to provide a basic overview of accretion disc outbursts and stability analysis along with highlighting the importance of considering the hierarchy of timescales involved.

## II. Time-Dependent Accretion Flow Models

Time-dependent accretion flow models consider the disc's time evolution, which helps us to study the generation of instabilities that lead to outbursts. The magnitude of the viscosity governs the disc flow's time dependence [8,11,40]. Thus, one of the few sources of quantitative data regarding disc viscosity is provided by observations of time-dependent disc models. These models helps us to pinpoint the origins of various instabilities, monitor their growth rates and determine their eventual impacts [2]. The time dependence study mainly deals with the phenomenon of dwarf nova outbursts and is a field of great interest on its own.

In traditional steady-state optically thick discs, the viscosity has little effect on the observable properties, which is fortunate for confirming their existence, but it also implies that it is unlikely that observations of steady discs will provide much information about the viscosity [2,5,43]. Moreover these models failed to explain the time evolution phenomenon and instabilities occurring in outbursting candidates in binary systems. This is where the need of time-dependent accretion flow models comes into picture. Instabilities in accretion discs are caused by a variety of physical mechanisms.

## III. The Hierarchy of Timescales

When studying instabilities in accretion discs, it's important to examine the hierarchy of timescales. The three main timescales are:

### A. Dynamical Timescale

The shortest timescale in an accretion disc. This timescale characterizes the disc material's orbital period around the central object. In other words, it is the amount of time, the disc particles require to move around the accreting object in a single Kepler orbit. In addition, it is the characteristic period for restoring hydrostatic equilibrium perpendicular to the disc plane [1,2,9,28].

### B. Thermal Timescale

The timescale on which the disc can adjust its temperature in response to external perturbations. The thermal time scale indicates the duration needed for viscous forces to produce the thermal energy of a specific disc annulus. It is calculated by dividing the total thermal energy present by the local energy dissipation rate [1,2,9,28].

### C. Viscous Timescale

The viscous time or radial drift timescale refers to the period required for the fluid in a disc to move substantially in a radial direction. We can also say that it is the period needed for matter to slowly move inward due to viscosity. The viscous timescale is substantially longer than the thermal timescale and the dynamical timescale is shorter than the later [1,2,9,28].

The accretion disc's stability depends on all three timescales. Radiative processes on the thermal timescale alter the disc's vertical structure, while viscous forces on the viscous timescale drive angular momentum transfer [7,10,40]. A comprehensive stability analysis must encompass all three timescales in a coherent framework to accurately understand the origin of possible instabilities resulting in outbursts [1,2,39].

## IV. INSTABILITIES IN THE ACCRETION DISC

Various instabilities can arise in accretion discs due to the interaction between these timescales. Some common instabilities include:

### A. Thermal instability

It is caused by temperature fluctuations within the disc. In a disc annulus, when rate of heating is out of step with the rate of cooling, then a non-equilibrium situation arises. In this case it is said that the disc is subjected to thermal instability. This instability grows in thermal timescale [1,2,7,30].

### B. Viscous instability

Viscous instabilities occur due to the buildup of viscous stress in the disc (resulting from changes in viscosity or angular momentum transfer). We can explain viscous instability by applying a perturbation in the form of extra mass to a disc ring. If the extra mass we added as a disturbance, diffuses or drifts away from the disc ring and and the disc ring gradually returns to its original surface density, the disc ring is said to be viscously stable. Otherwise, it is considered viscously unstable [1,2,30].

### C. Thermal-Viscous instability

Examining how an accretion disc reacts to perturbations, which might be local or global, is necessary to determine the stability of the disc. The most well-known and researched instability is the thermal-viscous instability, which is responsible for some X-ray binary outbursts as well as dwarf nova outbursts [29]. When angular momentum is transported viscously and there is a non-linear feedback between the heating and cooling processes, thermal-viscous instability results. Under some circumstances, a local region in the disc may change from a cool, low-viscosity state to a hot, high-viscosity one. The mass accretion rate significantly increases because of the change in the viscous regime, which in turn impacts the disc's heating and cooling processes. The hot state's temperature increase causes a higher equilibrium mass accretion rate, which enhances the outburst [26].

The basic principles of thermal-viscous instability can be explained by the S-shaped equilibrium curve [7] (Fig.1.). In an accretion disc, hydrogen is the most abundant element. The degree of ionization of hydrogen determines the stability of the disc [38,39,40]. The hydrogen fully ionized state is known as the high state or HII state. However, in low state or HI state, hydrogen is neutral. Both these branches are stable (thermally and viscously) [30,31]. Partially ionized hydrogen occurs in the intermediate state. In the accretion disc's outer regions, when the temperatures vary between $\log_{10}T = 3.5$–$4$ K, the partial ionization of hydrogen occurs [27,28]. During partial ionization of hydrogen, the disc may become unstable due to changes in opacity. In the context of viscous and thermal instabilities, this state is unstable [30]. Three solutions are found for a specific range of surface density $\Sigma$ ($\Sigma B < \Sigma < \Sigma A$) for an accretion disc annulus [30]. Let us consider a disc ring where $\dot{M}_{out}$ and $\dot{M}_{in}$ are mass outflow and inflow rate from and into the ring respectively. In the HI state, where $\dot{M}_{in} > \dot{M}_{out}$, surface density profile as well as disc mass will increase, until a maximum density $\Sigma A$ is reached and jumps to HII state [30]. In HII state, $\dot{M}_{in} < \dot{M}_{out}$ satisfies. This as a result decreases the disc's mass there. This continues until a minimum surface density $\Sigma B$ is reached and the disc annulus jumps to HI state. The outbursts phenomena result from the constant oscillations between these two states.

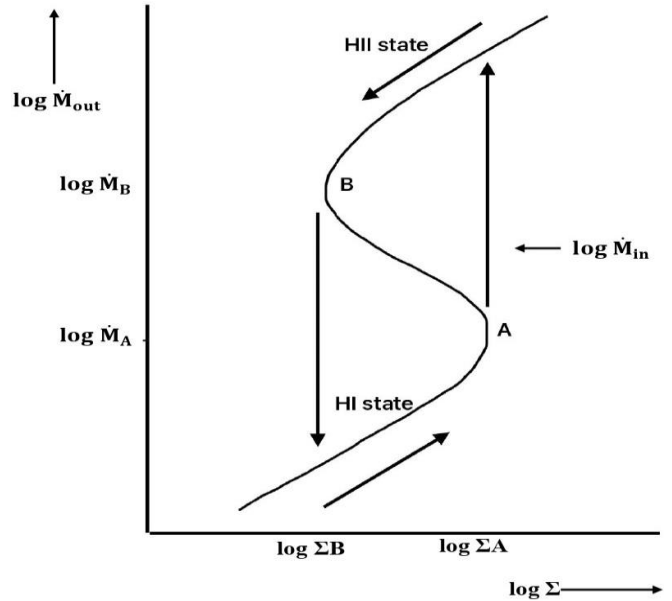

Fig.1. Schematic diagram of S-shaped equilibrium curve.

### D. Radiation Pressure Instability

When the pressure of radiation surpasses the pressure of gas, the black hole accretion disc, which has a classical heating component proportional to the pressure and viscosity parameter $\alpha$ (a parameter that describes the efficiency of angular momentum transport in an accretion disc) [20,42], can experience both thermal and viscous instabilities. This happens within the accretion disc's innermost radii that encircle a compact object (in case of black holes and neutron stars) [20,21]. Shakura and Sunyaev (1973) identified radiation pressure instability in their traditional $\alpha$-models early on [22,43], which they thoroughly examined in 1976. According to Taam et al. [24] and Deegan et al. [23], the microquasar GRS 1915+105's recurrent outbursts lasting hundreds of seconds could be explained by radiation pressure instability.

The magnetorotational instability (MRI), which is caused by the interplay between weak magnetic fields and the disc's differential rotation, is another significant instability [39]. Viscosity can be efficiently enhanced, and angular momentum can be transported within the disc via an MRI. On the other hand, stability analysis of discs under MRI influence is a more complex field that is still being investigated.

## V. STABILITY ANALYSIS

To analyze the stability of an accretion disc, we require knowledge on how the perturbations or disruption to its structure grows and evolves over time. Linear stability analysis is commonly used to examine how a system responds to small perturbations [1,42]. After the addition of perturbation, if the perturbation grows, it may indicate an instability that could lead to sudden and intense outburst of energy.

*A. Linear Stability Analysis*

Linear stability analysis involves simplifying the basic governing equations of the flow adding small perturbations to the disc's equilibrium state to examine their stability [1,19,34]. By solving the simplified equations, we can determine the conditions which give rise to instabilities. Fujimoto and Arai [18], investigated the stability of an optically thick, slim accretion disc around a black hole. They obtained a dispersion relation of the fourth order within the context of the study of linear stability. Also, for the cataclysmic variable disc models and their boundary layer (BL), Collins et al. [19] reported on the findings of a linearized perturbation analysis using local, linear stability analysis.

*B. Non-Linear stability analysis*

Although linear stability analysis helps us to know how perturbations first develop, nonlinear analysis and numerical simulations are mostly required to fully understand occurrence of instabilities. Non-linear effects can lead to a range of outcomes, which includes the formation of turbulent patterns [16,17], interactions between various unstable modes [14] and saturation of large-scale instabilities [15]. Balbus et al. [16] used a mix of analytical and numerical approach to explore the hydrodynamical non-linear stability of fluid flows that rotate differentially at different distances from the center (differential rotation) and flows that slide past each other (pure shear flows) in three-dimensions.

*C. Numerical Simulations*

Numerical simulations work as the most effective way to investigate the broad range dynamics of accretion discs [10,12,13,35]. Simulations can help us to understand how the interplay between thermal, viscous, and magnetorotational instabilities lead to disc outbursts and fluctuations [36,37,38]. Bergaulinger et al. [13] used semi-global simulations to explore the growth and saturation of the MRI in core collapse supernovae by studying its evolution. The accretion-ejection instability in magnetized accretion discs is simulated numerically by Caunt and Tagger [12].

## VI. OBSERVATIONAL IMPLICATIONS OF OUTBURSTS

Accretion disc outbursts can be observed in a variety of astrophysical systems, including cataclysmic variables (dwarf novae) and X-ray binaries.

*A. Cataclysmic variables*

This kind of binary system consists of a white dwarf which draws matter from a companion star [1,4,29]. Primarily, the companion star is a main sequence star or red dwarf [27]. Of all known cataclysmic variables, nearly fifty percent belong to the dwarf novae class, which has a known orbital period [1,41]. The outbursts last a few days, then happen again in a few weeks or months, and have an amplitude ranging from two to five magnitudes [1,25,43]. The recurrence time and the form of the outburst light curves are not exactly periodic. The most studied example is the SS Cygni system [6]. The spectral characteristics of dwarf novae in outbursts are very similar to those of nova-like variables that are continuously bright, or cataclysmic variables without dwarf nova outbursts. It is assumed that dwarf novae also feature a quasi-steady state disc during outbursts since nova-likes are believed to have bright, steady-state accretion discs [1].

*B. X-ray Binaries*

A compact object (a neutron star or black hole) that is accumulating mass from a companion star is what makes up an X-ray binary [1,2,4,5]. These systems are highly variable, exhibiting outbursts capable of several orders of magnitude increase in X-ray luminosity. Soft X-ray transients are a kind of low mass X-ray binaries that exhibit outbursts that are similar to dwarf novae outbursts, but with a considerably longer timescale and amplitude [1,3,32]. Compared to dwarf novae, the form of the light curve is significantly more variable. They have a typical shape which rises quickly over a few days and then slowly decays exponentially over several months. While there is concurrent brightening in other wavebands, particularly in the optical range, the outburst is most noticeable in the X-ray regime. There has only been one observation of several soft X-ray transient outbursts [1]. This suggests that the recurrence period is extremely long (decades, centuries, or even longer). Certain systems are seen repeating every year or every few years.

Studying accretion disc outbursts may enhance our understanding of other astrophysical systems, including active galactic nuclei (AGN) [20,21] and gamma-ray bursts (GRB) [33]. These systems show rapid and extreme brightness increases, possibly due to similar instabilities in their accretion discs.

## VII. CONCLUSION

This paper presents a summary of current state of knowledge about stability analysis and accretion disc outbursts. It is necessary to use time-dependent models which use hierarchy of dynamical, thermal, and viscous timescales to explain accretion disc outbursts. The various instabilities which lead to outbursts are also reviewed here. We have also discussed linear, non-linear and numerical stability analysis which are essential for determining the factors that contribute to outbursts. Investigation of accretion disc outbursts can help us to understand the importance of different complex binary systems along with other astrophysical systems which display similar behaviors. Further research in this area is necessary for fully understanding the process that drives outbursts in accretion discs.

ACKNOWLEDGMENT

The authors LD, AJB and BS acknowledge the reviewer for thoroughly reviewing the manuscript and providing beneficial comments. The authors also acknowledge the use of Meta AI and Google Gemini AI to improve the readability of the text.

## REFERENCES

[1] Kolb, U., *Extreme Environment Astrophysics*. 2010.

[2] Frank, J., King, A., and Raine, D. J., *Accretion Power in Astrophysics: Third Edition*. 2002, p. 398.

[3] Mineshige, S., Kim, S.-W., and Wheeler, J. C., "Time-dependent X-Ray Emission from Unstable Accretion Disks around Black Holes", *The Astrophysical Journal*, vol. 358, IOP, p. L5, 1990. doi:10.1086/185766.

[4] Seward, F. D. and Charles, P. A., *Exploring the X-ray Universe*. 2010.

[5] Courvoisier, T. J.-L., *High Energy Astrophysics: An Introduction*. 2013. doi:10.1007/978-3-642-30970-0.

[6] Shapiro, S. L. and Teukolsky, S. A., *Black holes, white dwarfs and neutron stars. The physics of compact objects*. 1983. doi:10.1002/9783527617661.

[7] Bambi, C., "Astrophysics of Black Holes", in *Astrophysics of Black Holes: From Fundamental Aspects to Latest Developments*, 2016, vol. 440. doi:10.1007/978-3-662-52859-4.

[8] Roy, A., "Outbursts in Stellar Black Hole Candidates: A Time-Dependent Study of Viscous Accretion Flow", in *Exploring the Universe: From Near Space to Extra-Galactic*, 2018, vol. 53, p. 145. doi:10.1007/978-3-319-94607-8_12.

[9] Skipper, C., *Fast Spectral Variability in the X-ray Emission of Accreting Black Holes*. 2015.

[10] Giri, K., *Numerical Simulation of Viscous Shocked Accretion Flows Around Black Holes*. 2015.

[11] Hameury, J.-M., Menou, K., Dubus, G., Lasota, J.-P., and Hure, J.-M., "Accretion disc outbursts: a new version of an old model", *Monthly Notices of the Royal Astronomical Society*, vol. 298, no. 4, OUP, pp. 1048–1060, 1998. doi:10.1046/j.1365-8711.1998.01773.x.

[12] Caunt, S. E. and Tagger, M., "Numerical simulations of the accretion-ejection instability in magnetised accretion disks", *Astronomy and Astrophysics*, vol. 367, pp. 1095–1111, 2001. doi:10.1051/0004-6361:20000535.

[13] Obergaulinger, M., Cerdá-Durán, P., Müller, E., and Aloy, M. A., "Semi-global simulations of the magneto-rotational instability in core collapse supernovae", *Astronomy and Astrophysics*, vol. 498, no. 1, pp. 241–271, 2009. doi:10.1051/0004-6361/200811323.

[14] Li, Z., Wang, X. Q., Xu, Y., Liu, H. F., and Huang, J., "Nonlinear interaction between double tearing mode and Kelvin-Helmholtz instability with different shear flows", *Scientific Reports*, vol. 13, 2023. doi:10.1038/s41598-023-40920-0.

[15] Tur, A., Chabane, M., and Yanovsky, V., "Saturation of a large scale instability and non linear stuctures in a rotating stratified flow", *arXiv e-prints* 2014. doi:10.48550/arXiv.1406.3962.

[16] Balbus, S. A., Hawley, J. F., and Stone, J. M., "Nonlinear Stability, Hydrodynamical Turbulence, and Transport in Disks", *The Astrophysical Journal*, vol. 467, IOP, p. 76, 1996. doi:10.1086/177585.

[17] Balbus, S. A. and Hawley, J. F., "Instability, Turbulence, and Enhanced Transport in Accretion Disks", in *IAU Colloq. 163: Accretion Phenomena and Related Outflows*, 1997, vol. 121, p. 90.

[18] Fujimoto, S. and Arai, K., "Stability of a slim accretion disk around a black hole", *Astronomy and Astrophysics*, vol. 330, pp. 1190–1196, 1998.

[19] Collins, T. J. B., Helfer, H. L., and van Horn, H. M., *Oscillations of Accretion Disks and Boundary Layers in Cataclysmic Variables*, vol. 194, 1999.

[20] Janiuk, A. and Czerny, B., "On different types of instabilities in black hole accretion discs: implications for X-ray binaries and active galactic nuclei", *Monthly Notices of the Royal Astronomical Society*, vol. 414, no. 3, OUP, pp. 2186–2194, 2011. doi:10.1111/j.1365-2966.2011.18544.x.

[21] Janiuk, A., Misra, R., Czerny, B., and Kunert-Bajraszewska, M., "Stability of black hole accretion disks", in *European Physical Journal Web of Conferences*, 2012, vol. 39. doi:10.1051/epjconf/20123906004.

[22] Lightman, A. P. and Eardley, D. M., "Black Holes in Binary Systems: Instability of Disk Accretion", *The Astrophysical Journal*, vol. 187, IOP, p. L1, 1974. doi:10.1086/181377.

[23] Deegan, P., Combet, C., and Wynn, G. A., "The outburst duration and duty cycle of GRS1915+105", *Monthly Notices of the Royal Astronomical Society*, vol. 400, no. 3, OUP, pp. 1337–1346, 2009. doi:10.1111/j.1365-2966.2009.15573.x.

[24] Taam, R. E., Chen, X., and Swank, J. H., "Rapid Bursts from GRS 1915+105 with RXTE", *The Astrophysical Journal*, vol. 485, no. 2, IOP, pp. L83–L86, 1997. doi:10.1086/310812.

[25] Cannizzo, J. K., Ghosh, P., and Wheeler, J. C., "Convective accretion disks and the onset of dwarf nova outbursts.", *The Astrophysical Journal*, vol. 260, IOP, pp. L83–L86, 1982. doi:10.1086/183875.

[26] Faulkner, J., Lin, D. N. C., and Papaloizou, J., "On the evolution of accretion disc flow in cataclysmic variables- I.The prospect of a limit cycle in dwarf nova systems.", *Monthly Notices of the Royal Astronomical Society*, vol. 205, OUP, pp. 359–375, 1983. doi:10.1093/mnras/205.2.359.

[27] Hōshi, R., "Accretion Model for Outbursts of Dwarf Nova", *Progress of Theoretical Physics*, vol. 61, no. 5, pp. 1307–1319, 1979. doi:10.1143/PTP.61.1307.

[28] Livio, M., "Accretion Discs: Limit Cycles and Instabilities", in *Astrophysical Discs - an EC Summer School*, 1999, vol. 160, p. 33. doi:10.48550/arXiv.astro-ph/9810035.

[29] Meyer, F. and Meyer-Hofmeister, E., "On the elusive cause of cataclysmic variable outbursts.", *Astronomy and Astrophysics*, vol. 104, pp. L10–L12, 1981.

[30] Mineshige, S., "Accretion Disk Instabilities", *Astrophysics and Space Science*, vol. 210, no. 1–2, Springer, pp. 83–103, 1993. doi:10.1007/BF00657876.

[31] Mineshige, S. and Osaki, Y., "Disk-instability model for outbursts of dwarf novae Time-dependent formulation and one-zone model", *Publications of the Astronomical Society of Japan*, vol. 35, no. 3, OUP, pp. 377–396, 1983.

[32] Mineshige, S. and Wheeler, J. C., "Disk-Instability Model for Soft X-Ray Transients Containing Black Holes", *The Astrophysical Journal* vol. 343, IOP, p. 241, 1989. doi:10.1086/167701.

[33] Janiuk, A., Yuan, Y., Perna, R., and Di Matteo, T., "Instabilities in the Time-Dependent Neutrino Disk in Gamma-Ray Bursts", *The Astrophysical Journal*, vol. 664, no. 2, IOP, pp. 1011–1025, 2007. doi:10.1086/518761.

[34] Latter, H. N., Fromang, S., and Faure, J., "Local and global aspects of the linear MRI in accretion discs", *Monthly Notices of the Royal Astronomical Society*, vol. 453, no. 3, OUP, pp. 3257–3268, 2015. doi:10.1093/mnras/stv1890.

[35] Hawley, J. F., "Numerical Simulations of MHD Accretion Disks", *Highlights of Astronomy*, vol. 15, pp. 237–238, 2010. doi:10.1017/S1743921310009014.

[36] Kadam, K., Vorobyov, E., Regály, Z., Kóspál, Á., and Ábrahám, P., "Outbursts in Global Protoplanetary Disk Simulations", *The Astrophysical Journal*, vol. 895, no. 1, IOP, 2020. doi:10.3847/1538-4357/ab8bd8.

[37] Ross, J., Latter, H. N., and Tehranchi, M., "MRI turbulence and thermal instability in accretion discs", *Monthly Notices of the Royal Astronomical Society*, vol. 468, no. 2, OUP, pp. 2401–2415, 2017. doi:10.1093/mnras/stx564.

[38] Habibi, A. and Abbassi, S., "Thermal Instability of Thin Accretion Disks in the Presence of Wind and a Toroidal Magnetic Field", *The Astrophysical Journal*, vol. 887, no. 2, IOP, 2019. doi:10.3847/1538-4357/ab5793.

[39] Balbus, S. A. and Hawley, J. F., "A Powerful Local Shear Instability in Weakly Magnetized Disks. I. Linear Analysis", *The Astrophysical Journal*, vol. 376, IOP, p. 214, 1991. doi:10.1086/170270.

[40] Pringle, J. E., "Accretion discs in astrophysics", *Annual Review of Astronomy and Astrophysics*, vol. 19, pp. 137–162, 1981. doi:10.1146/annurev.aa.19.090181.001033.

[41] Osaki, Y., "An Accretion Model for the Outbursts of U Geminorum Stars", *Publications of the Astronomical Society of Japan*, vol. 26, OUP, p. 429, 1974.

[42] Piran, T., "The role of viscosity and cooling mechanisms in the stability of accretion disks.", *The Astrophysical Journal* vol. 221, IOP, pp. 652–660, 1978. doi:10.1086/156069.

[43] Shakura, N. I. and Sunyaev, R. A., "Black holes in binary systems. Observational appearance.", *Astronomy and Astrophysics*, vol. 24, pp. 337–355, 1973.